\documentclass[a4paper,11pt]{article}
\usepackage{pos}
\usepackage{slashed}%

\title{Twenty years of $\Theta^+$}

\author*{Michał Praszałowicz}

\affiliation{Institute of Theoretical Physics, Jagiellonian University,\\
  Łojasiewicza 11, Kraków, Poland}

\emailAdd{michal.praszalowicz@uj.edu.pl}

\abstract{Twenty years ago, in 2003, two experimental groups, LEPS  and DIANA, announced the discovery of a light, narrow, exotic baryon with mass within the range of 1540 MeV,  which was later dubbed as $\Theta^+$. In this talk we recall the history of this discovery and its theoretical foundations. We also discuss possible future experiments that could determine  the existence of $\Theta^+$.}

\FullConference{Corfu Summer Institute 2023 "School and Workshops on Elementary Particle Physics and Gravity" (CORFU2023)\\
 23 April - 6 May , and 27 August - 1 October, 2023,
Corfu, Greece\\}

\begin{document}
\maketitle

\section{Introduction}
Twenty years ago, in 2003, two experimental groups, LEPS~\cite{LEPS:2003wug}  and DIANA~\cite{DIANA:2003uet}, announced the discovery
of a light, narrow, exotic baryon with mass of the order of $1540$~MeV, 
which was later dubbed as $\Theta^+$. Since the decay channel was $KN$ they concluded
that the observed resonance was the lightest member of the antidecuplet of exotic pentaquark baryons, 
namely a $ u u d d \bar{s}$ state.
These searches were motivated by chiral models, in particular by Ref.~\cite{Diakonov:1997mm} and references therein,
which almost two decades earlier, predicted light pentaquark $\overline{\bf 10}$ flavor
multiplet of positive parity. 

Soon other experiments presented the results of their analyses in search of $\Theta^+$, 
some of which confirmed the initial reports of LEPS and DIANA. A summary of the experimental results can be found in 
Refs.~\cite{Danilov:2008uxa,Liu:2014yva,Hyodo:2020czb} and more recently in Ref.~\cite{Amaryan:2022iij}. 
Obviously, none of these experiments, including LEPS and DIANA, 
has been designed to search for pentaquarks. People used data
collected for other purposes. Only later were dedicated experiments conducted, but with mixed results. In 2004 $\Theta^+$
paved its way to the Particle Data Group (PDG) listings~\cite{PDG2004} as a three star resonance, in 2005 its significance was reduced to two stars,
and in 2007 it was omitted from the summary table. As of 2008, it is no longer listed by the PDG~\cite{PDG2008}.

One of the peculiar features of $\Theta^+$, if it it exists, is its very small decay width. Indeed, in
2006, the BELLE collaboration reported search results for $\Theta^+$~\cite{Belle:2005thz}. No formation signal of the $\Theta^+$ 
baryon was observed, and an upper limit on the $\Theta^+$  width was estimated:
 $\Gamma < 0.64$~MeV for $M_{\Theta^+} = 1539$~MeV. Also DIANA in 2006 \cite{DIANA:2006ypd} confirmed their initial observation 
 with the mass of $M_{\Theta^+} = 1537 \pm 2$~MeV estimating the width: $\Gamma = 0.36\pm 0.11$~MeV. Is such a small width
 {\em unnatural}? -- a question often raised against $\Theta^+$. In our opinion: not really. Let us recall that
 recently the LHCb collaboration at CERN reported an excited $\Omega_c(3050)$ state with a reportedly very small
 width:
$\Gamma=0.8\pm 0.2 \pm 0.1$~MeV~\cite{Aaij:2017nav}.  
 In a later publication from 2021 \cite{LHCb:2021ptx} the LHCb collaboration concluded that: {\em  The natural width of the
$\Omega_c(3050)$ is consistent with zero}.
 $\Omega_c(3050)$ was found in the decay to $\Xi_c K^-$~\cite{Aaij:2017nav} where
 the kaon momentum is $p=275$~MeV. This is approximately only $\sim10$~MeV above the kaon momentum in the decay
 of $\Theta^+$. From this perspective the small pentaquark width is not particularly {\em unnatural}. As a consequence, the small width of $\Omega_c(3050)$
led to its interpretation as a heavy charm pentaquark belonging to the exotic SU(3) $\overline{\bf 15}$ 
multiplet~\cite{Kim:2017khv,Kim:2017jpx,Praszalowicz:2022hcp}.

Clearly, most non-observation experiments do not really exclude the existence of $\Theta^+$, but rather put an upper limit on its production cross-section.
The cleanest and decisive experiment would be the so called  {\em formation experiment} where the resonance is directly produced in $KN$ reaction.
DIANA is exactly such an experiment where the liquid Xenon bubble chamber was exposed to a separated $K^+$ beam. On the contrary, LEPS
was a photoproduction experiment on a carbon nucleus $^{12}C$. 
In the follow-up analyses both DIANA~\cite{DIANA:2006ypd,DIANA:2009rzq,DIANA:2013mhv}, and 
LEPS in the dedicated photoproduction experiment on deuteron~\cite{LEPS:2008ghm,Nakano:2010zz}, confirmed their initial findings.
These results have survived unchallenged to this day.

In this paper in Sect.~\ref{sec:ChiMs} we briefly describe chiral models, and in Sec.~\ref{sec:Baryons} the emergence of baryons including
exotica. Then, in Sec.~\ref{sec:phenomenology}, we discuss phenomenology including
$\Theta^+$ mass and width, and briefly the properties of other members of $\overline{\bf 10}$. Most important
experiments are reviewed in Sec.~\ref{sec:exps}. Summary is given in Sec.~\ref{sec:sum}.

\section{Chiral Models and Solitons}
\label{sec:ChiMs}

Chiral models are effective models for Quantum Chromodynamics (QCD), which explore chiral symmetry and its spontaneous
and explicit breaking, and are tractable in the low energy regime. 
One can imagine that we integrate out from the QCD Lagrangian  gluon fields.  We are then left with the quark degrees of freedom only, which
have the canonical 
kinetic and possibly mass term, however the interaction Lagrangian  consists of an infinite number of nonlocal many-quark vertices.
Nevertheless, this effective Lagrangian will
be chirally invariant. One can
truncate this Lagrangian to a local four-quark interaction only, the so called 
Nambu--Jona-Lasinio model~\cite{Nambu:1961tp,Nambu:1961fr}. 

To ensure chiral invariance
it is convenient to introduce eight auxiliary pseudo-Goldstone fields $\boldsymbol{\varphi}$ (pions, kaons and $\eta$)
in a form of a unitary SU(3) matrix (for models including other meson fields see Ref.~\cite{Diakonov:2013qta})
\begin{equation}
U=\exp\left(i  \frac{2\boldsymbol{\lambda}\cdot \boldsymbol{\varphi}}{F} \right) \, ,
\end{equation}
where $\boldsymbol{\lambda}$ are Gell-Mann matrices and $F$ is a pseudoscalar (pion) decay constant that in the present normalization
is equal to 186~MeV.

The simplest Lagrangian following from the above procedure, a chiral quark model Lagrangian,  is given by
\begin{equation}
{\cal L}_{\chi{\rm QM}}=\bar{\psi}\left( i \slashed{\partial}-m -M U^{\gamma_5} \right)\psi \, ,
\label{chiQM}
\end{equation}
where
\begin{equation}
U^{\gamma_5}= U \frac{1+\gamma^5}{2}+ U^{\dagger}  \frac{1-\gamma^5}{2}\, .
\end{equation}

This remarkably simple Lagrangian has been in fact derived \cite{Diakonov:1985eg,DiakonovMogilany}
in the mid eighties
from the instanton picture of the QCD vacuum \cite{Diakonov:1983hh,Shuryak:1983ni}. Here $M$ denotes
the {\em constituent} quark mass of the order of 350~MeV and $m$ is a {\em current } mass matrix.

Chiral symmetry corresponds  to the independent global SU(3) rotations of  left and right fermions:
\begin{equation}
\psi_L \rightarrow L \psi_L, ~~~ \psi_R \rightarrow R \psi_R \, .
\label{eq:ChiraLR}
\end{equation}
Transformations (\ref{eq:ChiraLR}) leave the interaction term invariant if
\begin{equation}
U \rightarrow LUR^{\dagger} \, ,
\end{equation}
which is  nothing else but a nonlinear realization of chiral symmetry \cite{Scherer:2002tk}. Vacum state
corresponding to $U=1$ (or $\boldsymbol{\varphi}=0$) breaks this SU$_{\rm L}$(3)$\otimes$ SU$_{\rm R}$(3)
symmetry to
the vector SU(3): $L=R$, and breaks the axial symmetry $L=R^{\dagger}$.

We can further integrate the quark fields (employing suitable regularization). Then the kinetic part for the Goldstone bosons appears
\cite{Diakonov:1983bny,Balog:1984upv,Praszalowicz:1989dh,RuizArriola:1991gc} and we end up with a Lagrangian given in terms of the Goldstone bosons
alone. This Lagrangian is organized as a power series in Goldstone boson momenta, {\em i.e.} in terms of $\partial_{\mu}U$.
Such lagragians are used for precision calculations in chiral perturbation theory \cite{Scherer:2002tk}.

The first term in $\partial_{\mu}U$ expansion, a quadratic term, is fully dictated by the chiral symmetry, and is known as the Weinberg
Lagrangian \cite{Weinberg:1978kz}. Higher order terms of known group structure have, however, free coefficients that are not constrained
by any symmetry and have to be extracted from experimental data. Obviously, once we have at our disposal a reliable Lagrangian
like (\ref{chiQM}), we can compute effective Goldstone boson Lagrangian to any order in $\partial_{\mu}U$. 

A simple,
truncated Lagrangian with four derivatives only was proposed by Skyrme \cite{Skyrme:1961vq,Skyrme:1962vh} in 1961
and later generalized by Witten \cite{Witten:1983tw,Witten:1983tx}
\begin{equation}
{\cal L}_{\rm Sk}  = \frac{F^{2}}{16} 
    {\rm Tr}(\partial_{\mu}U^{\dagger}\partial^{\mu}U) 
+  \frac{1}{32e^{2}} 
{\rm Tr}([\partial_{\mu}U\,U^{\dagger},\partial_{\nu}U\,U^{\dagger}]^{2}) +{\cal L}_{\rm m} \, .
\label{eq:LSkyrme}
\end{equation}
The first term in (\ref{eq:LSkyrme}) is the Weinberg Lagrangian, the second one is called the {\em Skyrme term}. 
Parameter $e$ can be inferred from the pion scattering and is of the order $e=4\div6$. Here ${\cal L}_{\rm m}$
stands for an explicit mass Lagrangian for Goldstone bosons, which breaks chiral symmetry.

While both models (\ref{chiQM}) and (\ref{eq:LSkyrme}) can be expanded in powers of $\boldsymbol{\varphi}$ 
generating {\em perturbative} Goldstone boson
interactions, they also admit nonperturbative solutions of finite mass and size, namely the  {\em solitons}.

Soliton solutions in both models correspond to a specific static form of the chiral field $U$ known as a
 {\em hedgehog} Ansatz
\begin{equation}
U(\boldsymbol{r})=\exp \left( i\,\boldsymbol{n}\cdot \boldsymbol{\tau}%
\,P(r)\right) 
\label{eq:hedg}
\end{equation}
where $\boldsymbol{n}=\boldsymbol{r}/r$ and $\boldsymbol{\tau}=\{\lambda_1,\lambda_2,\lambda_3\}$.
Function $P(r)$ has to vanish at infinity, so that $U \rightarrow 1$. 
Hedgehog Ansatz (\ref{eq:hedg}) has a very special property: any spacial rotation of the unit vector $\boldsymbol{n}$ can be undone
by an internal SU(2) (isospin) rotation acting on Pauli matrices $\boldsymbol{\tau}$. 
This property is called {\em hedgehog symmetry}. 

\begin{figure}[h]
\centering
\includegraphics[width=9cm]{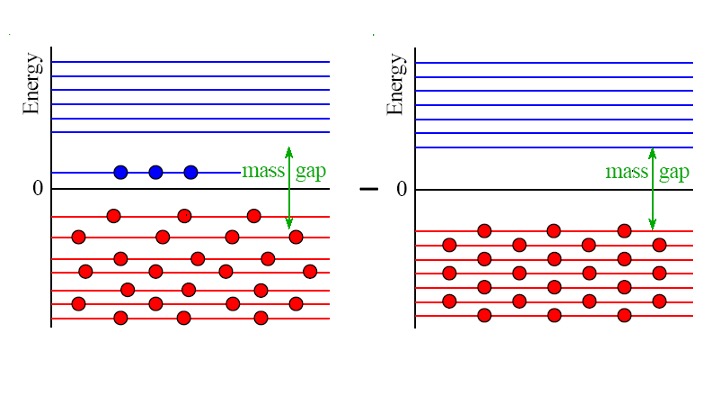} \vspace{-0.3cm} 
\caption{Schematic illustration of the calculation of the soliton mass  $\chi$QS model, which is the sum
over the energies of the valence quarks, and the properly regularized sum over the sea quarks with vacuum
contribution subtracted, see Eq.~(\ref{eq:Msol1}). }%
\label{fig:solitonlevels}%
\end{figure}

In a seminal paper from 1979  Witten suggested that that baryons emerge as solitons in the
chiral effective theory   \cite{Witten:1979kh}. While this is true both in the Nambu--Jona-Lasinio (NJL) model (\ref{chiQM}) and the Skyrme model
(\ref{eq:LSkyrme}), there are important differences. In the NJL (or Chiral Quark Soliton,  $\chi$QS for short) model, the specific form of the chiral background
field (\ref{eq:hedg}) leads to a rearrangement of the energy levels of the Dirac equation corresponding to (\ref{chiQM}). 
The lowest positive energy level falls into the mass gap, and the sea levels are distorted, leading to a stable static configuration corresponding to a self-consistently
determined form of the profile function $P(r)$ in (\ref{eq:hedg}). The energy of this configuration is computed 
as a regularized sum over all energy levels relative to the vacuum
(see {\em e.g.}~\cite{Goeke:2005fs}),
\begin{equation}
M_{\rm sol}=N_c\Big[ E_{\rm val}+\sum_{E_n<0} (E_n-E_n^{(0)})\Big].
\label{eq:Msol1}
\end{equation}
This is schematically illustrated in  Fig.~\ref{fig:solitonlevels}. Such a configuration carries no 
quantum numbers other than the baryon number following from the baryon number
of the valence quarks.

\begin{figure}[h]
\centering
\includegraphics[width=6cm]{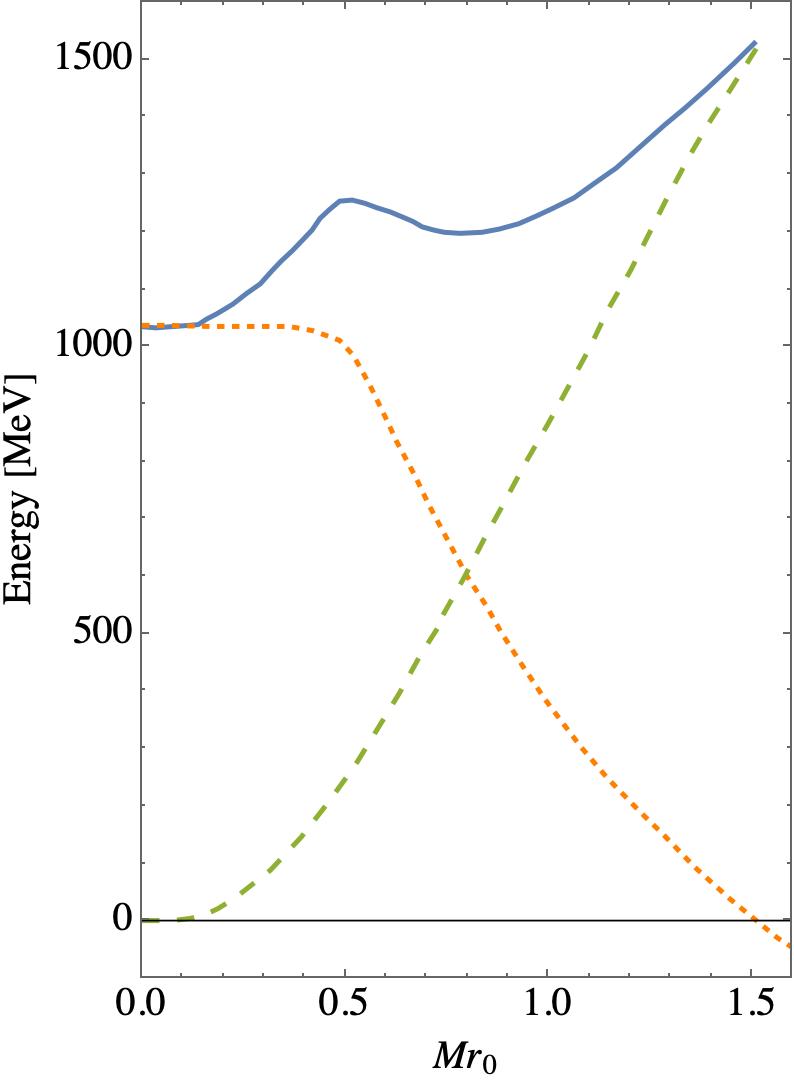} \vspace{-0.3cm}  
\caption{Soliton energy (mass) in MeV for $M=345$~MeV as a function of a dimensionless variational parameter $Mr_0$: solid (blue) -- total mass,
short dash (orange) -- energy of valence quarks, long dash (green) -- sea contribution. Minimum of $\sim1200$~MeV corresponds ro
$r_0 \simeq 0.5$~fm.
Figure from Ref.~\cite{Diakonov:1988mg}.}%
\label{fig:solitonenergy}%
\end{figure}

A convenient way of calculating the soliton mass (\ref{eq:Msol1}) is variational principle where the soliton size $r_0$
is treated as a variational parameter. The result is shown in Fig.~\ref{fig:solitonenergy}. We see that for large
soliton size   the valence level sinks into the Dirac sea, whereas for small soliton size entire energy is equal to
the energy of the valence quarks. These two limits are called the Skyrme Model (SM) limit and the Nonrelativistic Quark
Model (NRQM) limit, respectively. The $\chi$QS model interpolates between the two limits.

Therefore, the Skyrme Model can be viewed as a configuration shown in Fig.~\ref{fig:solitonlevels} where the valence level
sinked into the negative energy Dirac sea. In this case the soliton energy is given entirely by the energy of the distorted
 sea levels. An approximate formula for this energy can be obtained by a heat kernel expansion and can be expressed
in terms of functionals of the profile function $P(r)$ with no reference to the underlying Dirac structure. In the simplest
case it is just the energy corresponding to Lagrangian (\ref{eq:LSkyrme}) with the profile function $P(r)$ determined
from the pertinent equations of motion.

Let's recall that the hedgehog Ansatz $U(\boldsymbol{r})$ (\ref{eq:hedg}) can be viewed as a mapping of 
 $R^3 \rightarrow\,$SU(3), which is charcterized by a winding number $N_{\rm w}=P(0)/\pi$. According
 to Witten \cite{Witten:1979kh} $N_{\rm w}$ can be interpreted as a baryon number. In the case of the $\chi$QS model
 the condition $P(0)/\pi = {\rm integer}$ is not necessary~\cite{Diakonov:1988mg}, but the number of valence levels is given by $N_{\rm w}$.
 Moreover, the requirement that $P(0)/\pi = {\rm integer}$ ensures that the energy of the soliton in the Skyrme model
 is finite. Therefore solitons in the Skyrme model are referred to as {\em topological}, while in the NJL model
 as {\em nontopological}.
 
 So both models described above have soliton solutions of finite energy and size (corresponding to the {\em size} of the
 profile function $P(r)$), which have baryon number $B=1$ and no other quantum numbers. We call such a configuration a
 {\em classical} baryon. In the next Section we show how other quantum numbers such as isospin or spin are generated
 and discuss the following mass formulas.
 
 \section{Baryons in chiral models}
 \label{sec:Baryons}

In order to provide the {\em classical} baryon with specific
quantum numbers one has to consider an SU(3)-rotated time-dependent pseudoscalar
field
\begin{equation}
U(t,\boldsymbol{r})=A(t)U(\boldsymbol{r})A^{\dagger}(t)
\label{eq:rotU}
\end{equation}
and derive the pertinent Lagrangian expressed in terms of the collective velocities $da_{\alpha}(t)/dt$ defined as follows
\begin{equation}
A^{\dagger}(t) \frac{dA(t)}{dt} = \frac{i}{2}  \sum_{\alpha=1}^{8}
\lambda_{\alpha}\frac{da_{\alpha}(t)}{dt} \, .
\label{eq:adots}
\end{equation}

At this point it is important no note that
$A \in\,$SU(3)/U(1)  rather than full SU(3), since for the hedgehog Ansatz  (\ref{eq:hedg}) $ [\lambda_{8},U(\boldsymbol{r})] = 0 $.
 Therefore matrix $ A $ is defined up to a {\em local} U(1) factor
 $ h= \exp(i \lambda_{8} \phi) $, {\em  i.e.} $ A $ and $Ah$ are equivalent. For this reason the eighth coordinate $a_{8}(t)$
 is not dynamical and does not appear in the kinetic energy of the rotating soliton. Instead, it provides a constraint on the
 allowed states in the collective ({\em i.e.} corresponding to rotations (\ref{eq:rotU})) Hilbert space.
 
\begin{figure}[h]
\centering
\includegraphics[width=10cm]{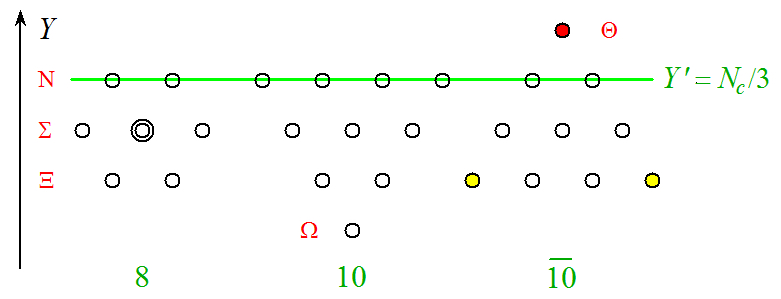} 
\caption{SU(3) representations selected by the constraint (\ref{eq:YR}). Isospin of states on the $Y'=1$ line
is equal to the soliton spin $J$.}%
\label{fig:irreps}%
\end{figure}

 Standard quantization procedure \cite{Guadagnini:1983uv,Jain:1984gp,Mazur:1984yf,Chemtob:1985ar} leads to the  rotational Hamiltonian,
 which has a form of a quantum mechanical symmetric top~\cite{Landau1977}
 \begin{equation}
{\cal{H}}_{\rm rot}= M_{\rm{sol}}
+ \frac{1}{2I_{1}}J(J+1) +\;\frac{1}{2I_{2}}\left[
C_{2}({\cal{R}})-J(J+1)-\frac{3}{4}Y^{\prime2}\right] 
\label{rotmass}%
\end{equation}
where $C_{2}(\mathcal{R})$ stands for the SU(3) Casimir operator and $J$ for the soliton angular momentum (spin). 
Soliton mass $ M_{\rm{sol}}$ and moments of inertia $I_{1,2}$ are calculable within a given model.
The constraint
mentioned above selects representations ${\cal R}$, which contain states of hypercharge $Y'$ equal to
\begin{equation}
Y' = \frac{N_{\rm c}}{3}\, . \label{eq:YR}
\end{equation}
Soliton spin $J$ is equal to the isospin of states on the $Y'$ rung of the pertinent weight diagram, Fig.\ref{fig:irreps}.

In the present case for
$N_c=3$ we have that $Y'=1$ and the allowed representations are 
\begin{equation}
{\cal R}= {\bf 8},\, {\bf 10}, \, \overline{ {\bf 10}}, \, {\bf 27}, \, {\bf 35}, \, \overline{ {\bf 35}}, \ldots \, .
\label{eq:SU3reps}
\end{equation}
We see that in addition to the octet and decuplet of positive-parity baryons, well known from the quark model, exotic representations, 
like $\overline{ {\bf 10}}$, emerge, all of positive parity.

Collective Hamiltonian and the constraint (\ref{eq:YR}) are exactly the same in the chiral quark model and in the Skyrme model.
The only obvious difference is that the soliton mass and the moments of inertia are in the Skyrme model expressed in terms of space
integrals over some functionals of the profile function $P(r)$, while in the case of the quark model they are given as regularized sums
over the one particle energy levels of the Dirac Hamiltonian corresponding to  Eq.~(\ref{chiQM}).

Wave functions for a quantum mechanical symmetric top are given in terms of Wigner $D^{({\cal R})}_{ab}$-functions~\cite{Landau1977}.
Skipping technicalities~\cite{Christov:1995vm}, baryon wave function takes the following form\footnote{One can find different representations of
this wave function in the literature that are equivalent to the one used here.}
\begin{equation}
\psi_{(B,\,J,J_{3})}^{(\mathcal{R})}(A)  
 = (-)^{J_{3}-Y^{\prime}/2}  \sqrt{\text{dim}(\mathcal{R})}\,
D_{(Y,\,T,\,T_{3}%
)(Y^{\prime},\,J,\,-J_{3})}^{(\mathcal{R})\ast}(A) \, .
\label{eq:rotwf}%
\end{equation}
Here $B=(Y,T,T_{3})$ stands for the SU(3) quantum numbers of a baryon in question,
and the second index of the $D$ function, $(Y^{\prime},J,-J_{3})$, corresponds
to the soliton spin. 

We see that the problem of baryon properties has been reduced to the quantum mechanical
Hamiltonian with well defined Hilbert space and explicit wave functions. One can therefore
compute all matrix elements needed for mass splittings, currents and other quantities.

For $m=0$ all states in a given representation ${\cal R}$ are degenerate. In order to compute
the mass splittings we have to express the symmetry breaking Hamiltonian, which is proportional to $m$, in terms of the collective
coordinates. In the $\chi$QS model the result reads
\begin{equation}
{\cal H}_{\mathrm{{br}}}=\alpha\,D_{88}^{(\boldsymbol{8})}(A)+\beta\,\hat{Y}+\frac{\gamma}{\sqrt{3}%
}\sum_{a=1}^{3}D_{8a}^{(\boldsymbol{8})}(A)\,\hat{J}_{a}, \label{eq:Hsb}%
\end{equation}
where $\alpha$, $\beta$, and $\gamma$ are proportional to the strange
quark mass.\footnote{For simplicity we assume $m_u=m_d=0$ and then $m=m_s$.} Furthermore,
$\alpha$  scales as $N_c$, and $\beta$ and $\gamma$
scale as $N_c^0$. $\hat{Y}$ and $\hat{J}_{a}$ are hypercharge and soliton spin operators, respectively.

Formally, $\beta$ and $\gamma$ are zero in the Skyrme model~\cite{Adkins:1983ya,Praszalowicz:1985bt}. 
In the large $N_c$ limit baryons consist from $N_c$ quarks, and therefore the hypercharge eigenvalue of the physical states
is also $Y\sim N_c$. This means that the second term in (\ref{eq:Hsb}), including $\hat{Y}$, is of the order 
${\cal O}(m_s N_c)$\footnote{Matrix elements of $D_{88}^{(\boldsymbol{8})}$ are ${\cal O}(N_c^0)$.} exactly as the first term
proportional to $\alpha$.
It was Gudagnini \cite{Guadagnini:1983uv} who argued that $\beta\,\hat{Y}$ should be included in the Skyrme model. 
In the chiral quark model it arises naturally
from $1/N_c$ expansion, similarly to the sub-leading term proportional to $\gamma$. Explicit expressions 
for the coefficients $\alpha,\,\beta$ and $\gamma$ are given {\em e.g.} in Eq.~(4) of Ref~\cite{Yang:2016qdz}
and the mass splitting including representation mixing are discussed in Ref.~\cite{Yang:2010fm}.

Since we have identified the symmetries of the soliton, it is straightforward to compute the pertinent currents, in particular the axial
current  \cite{Blotz:1994wi}. 
The axial current is of interest here, since via the Goldberger-Treiman relation it can be related to strong baryon decays.\footnote{This
approach to the width calculations has been criticised  in the literature, see {\em e.g.} Ref.~\cite{Walliser:2005pi}.}
In the non-relativistic limit for the initial and final baryons, $B_1$ and $B_2$ respectively,
the baryon-baryon-meson coupling can be written in the following form~\cite{Diakonov:1997mm}:
\begin{equation}
\mathcal{O}_{\varphi}=3 \sum_a \left[  G_{0}D_{\varphi\,a}^{(\boldsymbol{8})}-G_{1}\,d_{3ba}%
D_{\varphi\,b}^{(\boldsymbol{8})}\hat{J}_{a}-G_{2}\frac{1}{\sqrt{3}}D_{\varphi\,8}%
^{(\boldsymbol{8})}\hat{J}_{a}\right]  \,\frac{p_{i}}{M_{1}+M_{2}}  \label{eq:dec-op}%
\end{equation}
where $M_{1,2}$ denote masses of the
initial and final baryons and $p_{i}$ is the
c.m. momentum of the outgoing meson, denoted as $\varphi$, of mass 
$m$:%
\begin{equation}
\left\vert \boldsymbol{p}\, \right\vert =p=\frac{1}{2M_{1}}\sqrt{(M_{1}^{2}%
-(M_{2}+m)^{2})(M_{1}^{2}-(M_{2}-m)^{2})} \, .
\end{equation}
The factor of 3 in Eq.~(\ref{eq:dec-op}) is a matter of convenience.

The decay width is related to the matrix element of
$\mathcal{O}_{\varphi }$ squared, summed over the final and averaged
over the initial spin and isospin  denoted as $\overline{\left[ 
\ldots\right]  ^{2}}$,  see the Appendix of
Ref.~\cite{Diakonov:1997mm} for details of the corresponding
calculations:
\begin{equation}
\Gamma_{B_{1}\rightarrow B_{2}+\varphi}=\frac{1}{2\pi}\overline{\left\langle
B_{2}\left\vert \mathcal{O}_{\varphi}\right\vert B_{1}\right\rangle ^{2}%
}\,\frac{M_{2}}{M_{1}}p.
\label{eq:Gamma_phi}
\end{equation}
Factor $M_{2}/M_{1}$, used already in Ref.~\cite{Kim:2017khv}, is the same as in heavy baryon chiral perturbation
theory; see \emph{e.g.} Ref.~\cite{Cheng:2006dk,Cheng:2015dk1}. 

Here some remarks are in order. While the mass spectra are given as systematic
expansions both in $N_c$ and  $m_s$,  the decay widths cannot be organized in a similar way. 
They depend on modelling and ‘educated’ guesses, and hence are subject to additional
uncertainties \cite{Ellis:2004uz}. Most important uncertainty comes from the fact that the baryon masses $M_1$ and $M_2$
are formally infinite series in $N_c$ and $m_s$. 
The same holds for the momentum of the outgoing meson. It is a common practice
to treat the phase factor exactly, rather than expand it up to a given order in $N_c$
and $m_s$, despite the fact that in ${\cal O}_{\varphi}$ only a few first terms in $1/N_c$
and $m_s$ are included.  Here we have adopted
a convention with $M_1+M_2$ in (\ref{eq:dec-op}) and $M_2/M_1$ in (\ref{eq:Gamma_phi}), for other choice see {\em e.g.} \cite{Ellis:2004uz}.
Formally, in the large $N_c$ limit and small $m_s$ limit $M_1=M_2$ and all conventions are identical. Nevertheless, if we 
use physical masses for $M_{1,2}$, different conventions will result in different numerical results.

The leading term proportional to $G_0\sim N_c$ has been introduced already in the Skyrme model in Ref.~\cite{Adkins:1983ya},
whereas the subleading terms $G_{1,2}\sim N_c^0$ have been derived in the chiral quark model \cite{Diakonov:1997mm,Blotz:1994wi}.

Since we know the collective wave functions (\ref{eq:rotwf}), it is relatively straightforward to compute the matrix elements
for the mass splittings and decay widths. They are simply given in terms of the SU(3) Clebsch-Gordan coefficients~\cite{deSwart:1963pdg}.

\section{Antidecuplet phenomenology}
\label{sec:phenomenology}
\subsection{$\Theta^+$ mass}
\label{ssec:Thetamass}

In principle we can compute the classical mass, moments of inertia (\ref{rotmass}) and splitting parameters (\ref{eq:Hsb}) in 
either model (SM or $\chi$QS), however -- following Ref.~\cite{Adkins:1983ya} -- we adopt here a {\em model-independent} approach
and try to constrain these parameters from the data. Mass splittings between different multiplets are related to the moments of inertia
$I_{1,2}$ of the rotating soliton:
\begin{equation}
\Delta_{{\bf 10}-{\bf 8}}=\frac{3}{2} \frac{1}{I_1} \, ,~~~\Delta_{\overline{\bf 10}-{\bf 8}}=\frac{3}{2} \frac{1}{I_2}\, .
\label{eq:10-8split}
\end{equation}
Therefore, we have no handle on the strange moment of inertia $I_2$, and it is impossible to predict the mean 
antidecuplet mass from the non-exotic baryon masses alone.

Chiral symmetry breaking terms following from the fact that $m_s>m_{u,d} \simeq 0$ generate mass splittings within the SU(3)$_{\rm flavor}$ multiplets~\cite{Diakonov:1997mm}:
\begin{align}
\Delta M_{\bf 8}=&\frac{1}{20}\left(2\alpha+3\gamma\right) + \frac{1}{8}\,\Big[\left(2\alpha+3\gamma\right)+4\left(2\beta-\gamma\right) \Big] \, Y  \notag \\
 -&\frac{1}{20}\left(2\alpha+3\gamma\right) \Big[ T(T+1)-\frac{1}{4}Y^2 \Big] \, , \notag \\
\Delta M_{\bf 10}=&\frac{1}{16}\,\Big[\left(2\alpha+3\gamma\right)+8\left(2\beta-\gamma\right)\Big]\, Y\, ,\notag \\
\Delta M_{\overline{\bf 10}}=&\frac{1}{16}\,\Big[\left(2\alpha+3\gamma\right)+8\left(2\beta-\gamma\right)+4\gamma \Big]\,Y \, ,
\label{eq:massabg}
\end{align}
where parameters $\alpha,\, \beta$ and $\gamma$ are proportional to $m_s$. Note that in the Skyrme model $\gamma=0$, and $\beta=0$ if
we do not take into account the Guadagnini term~\cite{Guadagnini:1983uv}. Equations (\ref{eq:massabg}) are written in a form, 
from which one can immediately see that
mass splittings of non-exotic baryons depend in fact only on two combinations of parameters $\alpha,\, \beta$ and $\gamma$, namely
on $2\alpha+3\gamma$ and $2\beta-\gamma$, whereas
mass splittings in exotic $\overline{\bf 10}$ depend additionally on $\gamma$. 
Again, this means that we cannot predict mass splittings within $\overline{\bf 10}$ 
from the spectrum of non-exotic baryons.

Resorting to the model calculations one obtains a relatively small $\overline{\bf 10}-{\bf 8}$ splitting, namely $\Delta_{\overline{\bf 10}-{\bf 8}} \simeq 600$~MeV
\cite{Biedenharn:1984qg,Biedenharn:1984su}. One can now make a rough estimate of the $\Theta^+$ mass accepting the Skyrme model 
$\Delta_{\overline{\bf 10}-{\bf 8}}$ given above
 and assuming that the mass splittings in $\overline{\bf 10}$ are approximately equal to the ones in
the decuplet, $140 \div 150$~MeV. One obtains then that $\Theta^+$ mass is as low as $\sim1460$~MeV \cite{Praszalowicz:2003ik}. 
This is much lower than any quark model expectations.
More detailed analyses
in the Skyrme model \cite{Praszalowicz:2003ik,Praszalowicz:1987em} and in the quark-soliton model \cite{Diakonov:1997mm} 
led to the mass $1530 \div 1540$~MeV, which has been reinforced by the experimental
results of LEPS \cite{LEPS:2003wug} and DIANA \cite{DIANA:2003uet}.

\subsection{$\Theta^+$ width}
\label{ssec:Thetawidth}

There are two possible strategies to constrain the decay parameters $G_{0,1,2}$~(\ref{eq:dec-op}): one can either try to use directly data on strong decays,
or use the Goldberger-Treiman relation:
\begin{equation}
\left\{  G_{0},G_{1},G_{2}\right\}  =\frac{M_{1}+M_{2}}{2F_{\varphi}}\frac
{1}{3}\left\{  a_{0},-a_{1},-a_{2}\right\}
\label{eq:GTrelation}
\end{equation}
where constants $a_{0,1,2}$ enter the definition of the axial-vector
current~\cite{Blotz:1994wi,Praszalowicz:1998jm,paradox} and can be extracted from the
semileptonic decays of  the baryon octet~\cite{Yang:2015era}. 

Here, rather than discussing decay phenomenology in detail, we shall concentrate on a very peculiar
feature of the antidecupet decay constant (\ref{eq:Gamma_phi})
\begin{equation}
\overline{\left\langle N\left\vert \mathcal{O}_{K}\right\vert \Theta^+ \right\rangle ^{2}} \sim  G_{\overline{10}\rightarrow 8}^2,
~~~~~
G_{\overline{10}\rightarrow 8}=-a_0-\frac{N_c+1}{4}a_1-\frac{1}{2}a_2 \, ,
\end{equation}
where we have explicitly displayed the $N_c$ dependence following from the pertinent $N_c$ dependence of the flavor SU(3)
Clebsch-Gordan coefficients~\cite{Praszalowicz:2003tc}.

For small soliton size (the NRQM limit) one can compute constants $a_{0,1,2}$
analytically \cite{Praszalowicz:1998jm}
\begin{equation}
a_{0}\rightarrow-(N_{c}+2),\;a_{1}\rightarrow4,\;a_{2}\rightarrow2 \, .
\label{eq:NRa123}%
\end{equation}
which gives
\begin{equation}
G_{\overline{10}\rightarrow 8}=0 \, .
\label{eq:QMlimit4G}
\end{equation}
We see that the decay constant of antidecuplet is zero! The cancellation takes place for any $N_c$~\cite{Praszalowicz:2003tc}.
 This explains the smallness of the $\Theta^+$ width, which
for  realistic soliton size is not equal to zero, but still very small. In contrast, the decuplet 
decay constant is large, $G_{10\rightarrow8}=N_c+4$, explaining the large width of the $\Delta$ resonance. 

This is not the only reason why $\Theta^+$ width may be very small. The symmetry breaking Hamiltonian 
(\ref{eq:Hsb}) inevitably introduces representation mixing~\cite{Diakonov:1997mm,Praszalowicz:2004dn}, which
 in the case of the nucleon takes the following form
 \begin{equation}
 \mid N^{\rm phys} \rangle =\cos\alpha \mid N_{\bf 8} \rangle+ \sin\alpha \mid N_{\overline{\bf 10}} \rangle \, ,
 \label{eq:Nmixing}
 \end{equation}
where $\sin \alpha >0$ is small and therefore $\cos \alpha \simeq 1$. Note that $\Theta^+$ does not mix, and 
$\mid \Theta^{\rm phys} \rangle = \mid \Theta_{\overline{\bf 10}} \rangle$. Therefore, the decay of $\Theta^+$ to
$KN$ proceeds either directly to $\mid N_{\bf 8} \rangle$ or through mixing with $\mid N_{\overline{\bf 10}} \rangle$
\begin{equation}
g_{\Theta NK}\simeq G_{\overline{10}\rightarrow 8} +\sin \alpha \, H_{\overline{10}\rightarrow \overline{10}}
\label{eq:gThNKmix}
\end{equation}
leading to a new decay constant $H_{\overline{10}\rightarrow \overline{10}}$
\begin{equation}
H_{\overline{10}\rightarrow \overline{10}}=-a_0-\frac{5}{2} a_1 + \frac{1}{2} a_2 \, .
\end{equation}
From the model calculations and phenomenological studies one finds that $H_{\overline{10}\rightarrow \overline{10}}$ is large
and negative~\cite{Praszalowicz:2004dn,Goeke:2009ae,Praszalowicz:2010me}. Indeed, in the NRQM limit (\ref{eq:NRa123})
$H_{\overline{10}\rightarrow \overline{10}}=-4$.
This leads to a strong cancellation in Eq.~(\ref{eq:gThNKmix}), yielding the decay width of $\Theta^+$ very small.

The observation that the width of $\Theta^+$ may be quite small was perhaps the most important result of Ref.~\cite{Diakonov:1997mm}
since the smallness of its mass was anticipated a decade earlier~\cite{Biedenharn:1984qg,Biedenharn:1984su,Praszalowicz:1987em}
(see the discussion of the $\Theta^+$ width in Refs.~\cite{Jaffe:2004qj,Diakonov:2004ai,Jaffe:2004dc}).

\subsection{Other members of antidecuplet}

As explained after Eq.~(\ref{eq:massabg}), even if we anchor the exotic antidecuplet taking for the $\Theta^+$ mass 
the experimental value from LEPS and DIANA, we still
cannot predict the masses of other members of $\overline{\bf 10}$. Here we have two truly exotic pentaquarks corresponding 
to the corners of  antidecuplet (marked as yellow circles in Fig.~\ref{fig:irreps}), namely $\Xi^+$ and $\Xi^{--}$, 
whose quantum numbers cannot be constructed
from 3 quarks, and the remaining cryptoexotic states whose quantum numbers can be constructed both from 3 or 5 quarks.

In 2003 the NA49 Collaboration at CERN announced the observation of an exotic $\Xi^{--}$ pentaquark (lower left vertex in Fig.~\ref{fig:irreps}) at
1.862~GeV \cite{NA49:2003fxh}. If confirmed, it would be the second input besides $\Theta^+$ to anchor the exotic antidecuplet.
Unfortunately, 17 years later the successor of NA49 the NA61/SHINE Collaboration did not confirm the $\Xi^{--}$ peak around
1.8~GeV with 10 times greater statistics~\cite{NA61SHINE:2020mti}. One possible reason of this non-observation might be the  extremely
small width of $\Xi^{--}$. Indeed, in Ref.~\cite{Ellis:2004uz} it was argued that in the SU(3) symmetry limit ({\em i.e.} without mixing effects)
this width is up to a factor of $\sim 2$ equal to the width of $\Theta^+$, {\em i.e} of the order of 1~MeV. Original analysis of NA49~\cite{NA49:2003fxh} 
reported the width $\Xi^{--}$ below detector resolution of 18~MeV, while NA61/SHINE~\cite{NA61SHINE:2020mti} does not discuss
their sensitivity to the width of $\Xi^{--}$.

There exists, however, a potential candidate for a cryptoexotic pentaquark, namely the nucleon resonance $N(1685)$~\cite{Kuznetsov:2008ii}, which has been
initially announced  by GRAAL Collaboration at NSTAR Conference  in 2004 \cite{GRAAL:2004ndn}. $N(1685)$ has been observed 
in the quasi-free neutron cross-section and in the $\eta n$ invariant mass spectrum~\cite{GRAAL:2006gzl,Kuznetsov:2008hj} 
and was later confirmed by
CBELSA/TAPS \cite{CBELSA:2008epm} and LNS-Sendai \cite{Shimizu:2008cwq}. The observed structure 
can be interpreted as a narrow nucleon resonance with the mass 1685 MeV, total width $\le 25$~MeV and the photocoupling 
to the proton much smaller than to the neutron. Especially the latter property is easily understood assuming that $N(1985)$ is
a cryptoexotic member of $\overline{\boldsymbol{10}}$~\cite{Polyakov:2003dx,Yang:2013tka}.

The argument for small proton coupling is based on an approximate $U$-spin sub-symmetry of flavor SU(3). Both $\eta$ and photon are 
$U$-spin singlets and neutron and proton are $U$-spin triplet and doublet, respectively. The neutron- and proton-like members of
$\overline{\bf 10}$ are $U$-spin triplet and 3/2 multiplet, respectively. Therefore in the SU(3) symmetry limit proton photo-excitation
to $p_{\overline{\bf 10}}+\eta$ is forbidden, while neutron transition to $n_{\overline{\bf 10}}+\eta$ is allowed. For alternative explanations
see Refs.~\cite{Shklyar:2006xw,Anisovich:2008wd,Doring:2009qr}.

It was found that the width of $N(1685)$
is in the range of tens of MeV with a very small $\pi N$ partial width of
$\Gamma_{\pi N}\leq 0.5$~MeV \cite{str}. One should stress that the
decay to $\pi N$ is not suppressed in the SU$(3)$ limit and it
can be made small only if the symmetry violation is taken into account. Therefore in Ref.~\cite{Goeke:2009ae}
masses and widths of exotic $\overline{\bf 10}$ were reanalyzed taking into account mixing of the ground state
octet with antidecupelt, already discussed in Sect.~\ref{ssec:Thetawidth}, 
and anidecuplet mixing with the excited Roper resonance octet. Taking into account
all the available data on different branching ratios and some model input it was possible to constrain the mixing
angles\footnote{Due to the accidental equality of the SU(3) Clebsch-Gordan coefficients mixing angles of
$\Sigma$ and $N$ states in ocet and antidecuplet are equal, so only two mixing angles were necessary for
the discussed mixing pattern.} leading to 
\begin{align}
1795\;\text{MeV}  &  <M_{\Sigma_{\overline{10}}}<1830\;\text{MeV} \, ,%
\notag\\
1900\;\text{MeV}  &  <M_{\Xi_{\overline{10}}}<1970\;\text{MeV} \, 
\label{xirange}%
\end{align}
with decay widths
\begin{align}
9.7\;\text{MeV}  &  <\Gamma_{\Sigma_{\overline{10}}}<26.9\;\text{MeV} \, ,%
\notag\\
7.7\;\text{MeV}  &  <\Gamma_{\Xi_{\overline{10}}}<11.7\;\text{MeV} \, .
\label{xiGamma}%
\end{align}
These limits follow from the assumptions that $\Theta^+$ mass is 1540~MeV and its width is 1~MeV, and 
that the decay width of $N(1685)$ is smaller than 25~MeV. One sees that the decay width of $\Xi_{\overline{10}}$
is still small, but larger than in the SU(3) limit. Its mass is still in the range scanned by NA61/SHINE.

\section{Experiments}
\label{sec:exps}

The positive evidence for $\Theta^+$ by LEPS~\cite{LEPS:2003wug} and DIANA~\cite{DIANA:2003uet} 
has prompted a number of searches by other experimental groups. 
At that time, only data collected originally for searches other than $\Theta^+$ was available. Only later were dedicated experiments designed 
and conducted. For a complete list of experiments we refer the reader to reviews from 2008 \cite{Danilov:2008uxa},  from 2014 \cite{Liu:2014yva} and 
to a more recent review from 2022 \cite{Amaryan:2022iij}. Below we will briefly recall only a few experiments, mainly those that 
have so far uphold their initial positive results.

\subsection{Photoproduction}

Photoproduction on a nucleon  is not the best experiment to discover $\Theta^+$. Indeed, photon has to dissociate into an $s\bar{s}$ pair, and
the antistrange quark has to be injected into a nucleon. In practice $\gamma$ can dissociate in $K^+ K^-$ but not into $K^0 \bar{K}^0$. In the first
case $K^+$ may excite neutron to $\Theta^+$, but for $\Theta^+$ to be produced on a proton, $K^0$  must be replaced
by $K^{0\,*}$. Furthermore, since $g_{\Theta N K}$ is very small the  cross-section will be small as well. The situation is even worse for the
photoproduction on a proton, since $g_{\Theta N K^*}$ is not known. Estimates based on the SU(3) symmetry and experimental results on
$\eta$ photoproducion off the neutron show that the cross-section for the reaction $\gamma p \rightarrow \bar{K}^0 \Theta^+$ is of the order
of 1~nb \cite{Azimov:2006he}.  For this reason, the negative result of the CLAS experiment with the proton target in 2006 \cite{CLAS:2006anj}
 was not surprising (superseding earlier positive report  \cite{CLAS:2003wfm} from 2003).

Nevertheless, at LEPS (Laser-Electron Photon facility at SPring-8 in Japan) $\Theta^+$  peak was observed just
in photoproduction on a neutron inside a carbon nucleus:
$\gamma  n \rightarrow K^- \Theta^+$ and subsequently $\Theta^+ \rightarrow  K^+   n'$. Only kaons were detected, and the peak was observed in
  the missing mass $M_{\gamma  K^-}=E_{\gamma}-E_{K^-}$ distribution. The main problem was, however, that since the target neutron 
  was inside the carbon nucleus, its momentum was smeared
by  Fermi motion. After applying the Fermi motion correction, the $\Theta^+$ peak was clearly visible at $M_{\Theta^+}=1.54\pm0.01$~GeV 
with 4.6 $\sigma$ Gaussian significance. The width has been estimated to be smaller than 25~MeV.

Five years later, in 2008, LEPS published results from a dedicated photo-production experiment, this time on a deuteron target~\cite{LEPS:2008ghm}.
Although the measurement strategy was basically the same as in the case of carbon, the deuteron setup offered a possibility to cross-check the 
pentaquark production in a reaction $\gamma n \rightarrow K^- \Theta^+$ with $\Lambda(1520)$ production in $\gamma p \rightarrow K^+ \Lambda^0(1520)$.
This was possible because the LEPS detector has a symmetric acceptance for positive and negative particles. 
The analysis confirmed the existence of a narrow $\Theta^+$ signal
at $M_{\Theta^+}=1.524\pm0.002 \pm 0.003$~GeV. The significance has been estimated to be 5.1~$\sigma$ and the width much smaller than $30$~MeV.

In the meantime a number of experiments reported negative results, and skepticism about the existence of $\Theta^+$ was growing. Most importantly,
in the analogous experiment carried out by the CLAS (CEBAF
Large Acceptance Spectrometer\footnote{CEBAF stands for Continuous Electron Beam Accelerator Facility at Jefferson Laboratory
located in Newport News, VA, USA.}) 
collaboration,
no narrow peak corresponding to $\Theta^+$ was observed~\cite{CLAS:2006czw}, contradicting the earlier CLAS report from 2003~\cite{CLAS:2003yuj}.

The CLAS experiment is analogous to LEPS, but not identical. In $\gamma  d$ reaction
 CLAS observed all charged particles in the final state, including the spectator
proton. This required an elastic  rescattering of $K^-$ off the proton from a deuteron, so that the proton could acquire sufficient momentum to enable detection. 
The probability of such a rescattering was an essential factor in the CLAS analysis. Since LEPS assumed the proton to be a spectator,
the kinematic conditions of the two experiments were different. Moreover, the angular coverage of both detectors is also different:
less than 20 degrees for  LEPS and greater than 20 degrees for  CLAS in the LAB
system~\cite{Nakano:2017zrr}. 

To clarify the situation, the LEPS collaboration performed the search for $\Theta^+$  in $\gamma d \rightarrow K^+ K^- n p$ reaction
with 2.6 times higher statistics. The peak was still there.
In 2013--2014 a new measurement was performed with the improved proton acceptance. Partial results have been published
in different conference proceedings~\cite{Nakano:2017zrr,Nakano:2017fui,Yosoi:2019mno} but to the best of our knowledge, a full fledged journal article
has not yet been released. 

At the end of 2022, the LEPS2 detector started to collect new data in the search for $\Theta^+$ \cite{Nakanoprivate}. LEPS2 detector has
better angular coverage than LEPS and will look for $\Theta^+$ in the following  reactions~\cite{Ahn:2023hiu}: (1) 
$\gamma n \rightarrow K^- \Theta^+$ and (2) $\gamma p \rightarrow {\bar{K}}^{0\,*} \Theta^+$, where $\Theta^+$ will be reconstructed from
the following reactions  $\Theta^+ \rightarrow p K^{0}_{\rm S}\rightarrow p\, \pi^+\pi^-$ and in the second case additionally 
${\bar{K}}^{0\,*}\rightarrow K^- \pi^+$. Apparently, all four or five particles in the final state will be identified, which means that the uncertainty
of the previous measurements due the Fermi motion of the target  neutron or proton will be removed. 
Therefore, we look forward to future results.

To circumvent the problem of smal $g_{\Theta NK}$ coupling the authors of Ref.~\cite{Amarian:2006xt} proposed in 2006 to look for $\Theta^+$
at CLAS in the interference with $\phi$ meson, which is copiously produced. The interference cross-section is linear in the $\Theta^+$ coupling
and hence can be substantially larger than the production cross-section where  $\phi$ contribution is removed by kinematical cuts. Such analysis
was published six years later \cite{Amaryan:2011qc} with positive result. Nevertheless, this paper has not been formally approved be the
entire CLAS Collaboration, which criticised kinematical cuts applied in \cite{Amaryan:2011qc} and published an official disclaimer \cite{CLAS:2012gcw}.

\subsection{Resonance formation}

Unlike photoproduction, resonance formation in $KN$ scattering
is the cleanest experiment  possible in the search for $\Theta^+$. The Breit-Wigner
cross-section for a production of a resonance of spin $J$ and mass $M$ in the scattering of two hadrons of spin $s_1$ and $s_2$
takes the following form (see {\em e.g.} Eq.(51.1) in Ref.~ \cite{Workman:2022ynf})
\begin{equation}
\sigma_{\rm BW}(E)=\frac{2J+1}{(2s_1+1)(2s_2+1)}\frac{\pi}{k^2}B_{\rm in} B_{\rm out}\frac{\Gamma^2}{(E-M)^2+\Gamma^2/4},
\end{equation}
where $E$ is the c.m. energy, $k$ is the c.m. momentum of initial state and $\Gamma$ is the full width
at half maximum height of the resonance. The branching fraction for the resonance into the initial-state channel is $B_{\rm in}$
and into the final-state channel is $B_{\rm out}$ -- in the present case for $KN$ scattering and one of the possible final
states $K^+n$ or $K^0 p$ we have   $B_{\rm in} =B_{\rm out}=1/2$. Substituting the $\Theta^+$ mass one gets
that the cross-section at the peak $\sigma_{\rm BW}(M_{\Theta^+})\sim 15\div 20$~mb. This is a model independent prediction,
and we see that the  cross-section for $\Theta^+$ production in $KN$ scattering is large. More detailed study of 
$\Theta^+$ production in $K^+d \rightarrow K^0pp$ reaction shows that the production cross-section is in this
case of the order of 5~mb \cite{Sekihara:2019cot}. The feasibility study of searching for $\Theta^+$ in this channel at J-PARC has been
recently performed in Ref.~\cite{Ahn:2023hiu}.

The formation process was used in the DIANA experiment where the bubble chamber DIANA filled with liquid Xenon 
has been exposed to a  $K^+$ beam  from the ITEP proton synchrotron. In Ref.~\cite{DIANA:2003uet} the authors
analyzed the $K^0 p$ effective mass spectrum in the reaction $K^+n \rightarrow K^0 p$ on a nucleon bound in a Xenon
nucleus. A resonant enhancement with $M = 1539 \pm 2$~MeV and $\Gamma \le 9$~MeV was
observed. The statistical significance of the enhancement  was estimated to be 4.4~$\sigma$.

The DIANA collaboration continued analysis of the bubble chamber films and in 2006 published new results from the
larger statistics sample~\cite{DIANA:2006ypd}. They confirmed their initial observation with the mass of $M = 1537 \pm 2$~MeV with,
however,
much smaller estimate of the width: $\Gamma = 0.36\pm 0.11$~MeV. Depending
on the significance estimator they obtained statistical significance of 4.3, 5.3 or 7.3~$\sigma$.
Three years later they increased again statistics confirming the existence of $\Theta^+$ with approximately
the same mass and width, but higher statistical significance reaching 8~$\sigma$~\cite{DIANA:2009rzq}. These results
were confirmed in their last publication from 2014~\cite{DIANA:2013mhv}.

As already mentioned the formation experiment with $K^+$ beam could be easily performed at the J-PARC facility in Japan
looking at the three body final state $K^0 p p$~\cite{Ahn:2023hiu}.
Another very promising search for $\Theta^+$ will be possible at already approved program at $K_L$ facility  at JLab
\cite{Amaryan:2022iij,KLF:2020gai,Amaryan:2024koq}. Here, with a secondary beam of kaons, one may look at
a two-body reaction $K^0_L p \rightarrow K^+ n $ on the hydrogen target. The plan is to measure
the initial energy  benefiting from the design momentum resolution below 1 MeV, rather
than the invariant mass of $K^+n$ system. 
According to the current schedule data collection will start in 2026~\cite{Amaryan:2024koq}.
Note that the two-body final state is much cleaner than the three-body one,
which is proposed to be studied at J-PARC. Finally, at the $K_L$ facility one will  also 
be able to look for other members of antidecuplet, like $\Xi^+$.

\section{Summary}
\label{sec:sum}

Although  pentaquarks with heavy quarks are by now well established exotic baryons, $\Theta^+$ still remains elusive.
Its story is full of unexpected twists, emotions and lost hopes. The purpose of this review was to recall the theoretical 
basis of $\Theta^+$ and some experimental evidence. While the small mass of exotic antidecuplet can be easily
justified in the chiral models, where the $q\bar{s}$ pair is injected into a nucleon not as two independent quarks,
but rather as an almost massless Goldstone boson, its very small width has often been considered  {\em unnatural}.
In the Chiral Quark Soliton model the small width of $\Theta^+$  is {\em natural}  due to the cancellation of the
corresponding couplings in the non-relativistic limit. Unfortunately, since there is no intuitive argument 
as to why this is the case, many authors did not take it seriously. Moreover, the chiral models discussed here are largely 
considered only qualitative, although they are able to describe non-exotic baryons quite well. However, this is also true 
of other models that have trouble accommodating light exotic baryons.

The fate of $\Theta^+$ can only be decided experimentally. In our opinion the most promising are formation experiments
that can be conducted at the JLab and J-PARC facilities. The time scale is here of the order of a few years. Even earlier 
one may expect results from the photoporduction experiment at LEPS2. Whatever the results, we will eventually get a clear 
answer to the main question that has been bothering us in this article, whether $\Theta^+$ exists or not.

\newpage

\section*{Ackowledgements}

We would like to thank the organizers of the Corfu Workshop 2023 for their warm hospitality and 
the opportunity to present these thoughts. We would also like to thank Moskov Amaryan and Igor Strakovsky 
for their helpful comments and encouragement and Hyun-Chul Kim for long collaboration on exotica.


\begin{thebibliography}{99}
\bibitem{LEPS:2003wug}
T.~Nakano \textit{et al.} [LEPS],
{\em Evidence for a narrow S = +1 baryon resonance in photoproduction from the neutron},
Phys. Rev. Lett. \textbf{91} (2003) 012002,
doi:10.1103/PhysRevLett.91.012002,
[arXiv:hep-ex/0301020 [hep-ex]].

\bibitem{DIANA:2003uet}
V.~V.~Barmin \textit{et al.} [DIANA],
{\em Observation of a baryon resonance with positive strangeness in K+ collisions with Xe nuclei},
Phys. Atom. Nucl. \textbf{66} (2003)  1715,
doi:10.1134/1.1611587,
[arXiv:hep-ex/0304040 [hep-ex]].

\bibitem{Diakonov:1997mm}
D.~Diakonov, V.~Petrov and M.~V.~Polyakov,
{\em Exotic anti-decuplet of baryons: Prediction from chiral solitons},
Z. Phys. A \textbf{359} (1997)  305,
doi:10.1007/s002180050406,
[arXiv:hep-ph/9703373 [hep-ph]].


\bibitem{Danilov:2008uxa}
M.~Danilov and R.~Mizuk,
{\em Experimental review on pentaquarks},
Phys. Atom. Nucl. \textbf{71} (2008)  605,
doi:10.1134/S1063778808040029,
[arXiv:0704.3531 [hep-ex]].

\bibitem{Liu:2014yva}
T.~Liu, Y.~Mao and B.~Q.~Ma,
{\em Present status on experimental search for pentaquarks},
Int. J. Mod. Phys. A \textbf{29} (2014) 1430020,
doi:10.1142/S0217751X14300208,
[arXiv:1403.4455 [hep-ex]].


\bibitem{Hyodo:2020czb}
T.~Hyodo and M.~Niiyama,
{\em QCD and the strange baryon spectrum},
Prog. Part. Nucl. Phys. \textbf{120} (2021) 103868,
doi:10.1016/j.ppnp.2021.103868
[arXiv:2010.07592 [hep-ph]].

\bibitem{Amaryan:2022iij}
M.~Amaryan,
{\em History and geography of light pentaquark searches: challenges and pitfalls},
Eur. Phys. J. Plus \textbf{137}  (2022) 684,
doi:10.1140/epjp/s13360-022-02888-0,
[arXiv:2201.04885 [hep-ex]].

\bibitem{PDG2004}
S. Eidelman et al. (Particle Data Group), Phys. Lett. B 592 (2004) 1.

\bibitem{PDG2008}
C. Amsler et al. (Particle Data Group), Physics Letters B667 (2008) 1.

\bibitem{Belle:2005thz}
K.~Abe \textit{et al.} [Belle],
{\em Search for the Theta(1540)+ pentaquark using kaon secondary interactions at BELLE},
Phys. Lett. B \textbf{632} (2006)  173,
doi:10.1016/j.physletb.2005.10.077,
[arXiv:hep-ex/0507014 [hep-ex]].

\bibitem{DIANA:2006ypd}
V.~V.~Barmin \textit{et al.} [DIANA],
{\em Further evidence for formation of a narrow baryon resonance with positive strangeness in K+ collisions with Xe nuclei},
Phys. Atom. Nucl. \textbf{70} (2007) 35,
doi:10.1134/S106377880701005X,
[arXiv:hep-ex/0603017 [hep-ex]].

\bibitem{Aaij:2017nav} 
  R.~Aaij {\it et al.} [LHCb Collaboration],
  {\em Observation of five new narrow $\Omega_c^0$ states decaying to $\Xi_c^+ K^-$},
  Phys.\ Rev.\ Lett.\  {\bf 118} (2017) 182001,
doi:10.1103/PhysRevLett.118.182001,
[arXiv:1703.04639 [hep-ex]].
  
\bibitem{LHCb:2021ptx}
R.~Aaij \textit{et al.} [LHCb],
{\em Observation of excited $\Omega_c^0$ baryons in $\Omega_b^- \to \Xi_c^+ K^-\pi^-$decays},
Phys. Rev. D \textbf{104} (2021) 9,
doi:10.1103/PhysRevD.104.L091102,
[arXiv:2107.03419 [hep-ex]].

\bibitem{Kim:2017khv}
H.~C.~Kim, M.~V.~Polyakov, M.~Praszalowicz and G.~S.~Yang,
{\em Strong decays of exotic and nonexotic heavy baryons in the chiral quark-soliton model},
Phys. Rev. D \textbf{96} (2017) 094021,
[erratum: Phys. Rev. D \textbf{97}, no.3, 039901 (2018)],
doi:10.1103/PhysRevD.96.094021
[arXiv:1709.04927 [hep-ph]].
\bibitem{Kim:2017jpx}
H.~C.~Kim, M.~V.~Polyakov and M.~Prasza\l{}owicz,
{\em Possibility of the existence of charmed exotica},
Phys. Rev. D \textbf{96}  (2017) 014009,
doi:10.1103/PhysRevD.96.014009,
[arXiv:1704.04082 [hep-ph]].

\bibitem{Praszalowicz:2022hcp}
M.~Prasza\l{}owicz and M.~Kucab,
{\em Invisible charm exotica},
Phys. Rev. D \textbf{107} (2023) 034011,
doi:10.1103/PhysRevD.107.034011,
[arXiv:2211.01470 [hep-ph]].

\bibitem{DIANA:2009rzq}
V.~V.~Barmin \textit{et al.} [DIANA],
{\em Formation of a narrow baryon resonance with positive strangeness in K+ collisions with Xe nuclei},
Phys. Atom. Nucl. \textbf{73} (2010) 1168,
doi:10.1134/S1063778810070100,
[arXiv:0909.4183 [hep-ex]].

\bibitem{DIANA:2013mhv}
V.~V.~Barmin \textit{et al.} [DIANA],
{\em Observation of a narrow baryon resonance with positive strangeness formed in $K^+$Xe collisions},
Phys. Rev. C \textbf{89}  (2014) 045204,
doi:10.1103/PhysRevC.89.045204,
[arXiv:1307.1653 [nucl-ex]].



\bibitem{LEPS:2008ghm}
T.~Nakano \textit{et al.} [LEPS],
{\em Evidence of the $\Theta^+$ in the gamma d ---\ensuremath{>} K+ K- pn reaction},
Phys. Rev. C \textbf{79} (2009) 025210 ,
doi:10.1103/PhysRevC.79.025210,
[arXiv:0812.1035 [nucl-ex]].

\bibitem{Nakano:2010zz}
T.~Nakano [LEPS],
{\em Status of the Theta+ analysis at LEPS},
Nucl. Phys. A \textbf{835} (2010) 254,
doi:10.1016/j.nuclphysa.2010.01.200.

\bibitem{Nambu:1961tp}
Y.~Nambu and G.~Jona-Lasinio,
{\em Dynamical Model of Elementary Particles Based on an Analogy with Superconductivity. I.},
Phys. Rev. \textbf{122} 358 (1961) 345,
doi:10.1103/PhysRev.122.345.

\bibitem{Nambu:1961fr}
Y.~Nambu and G.~Jona-Lasinio,
{\em Dynamical Model of Elementary Particles Based on an Analogy with Superconductivity. II.},
Phys. Rev. \textbf{124} 254 (1961) 246,
doi:10.1103/PhysRev.124.246.

\bibitem{Diakonov:2013qta}
D.~Diakonov, V.~Petrov and A.~A.~Vladimirov,
{\em A theory of baryon resonances at large $N_c$},
Phys. Rev. D \textbf{88}  (2013) 074030,
doi:10.1103/PhysRevD.88.074030,
[arXiv:1308.0947 [hep-ph]].


\bibitem{Diakonov:1985eg}
D.~Diakonov and V.~Y.~Petrov,
{\em A Theory of Light Quarks in the Instanton Vacuum},
Nucl. Phys. B \textbf{272} (1986) 457,
doi:10.1016/0550-3213(86)90011-8.

\bibitem{DiakonovMogilany}
D.~Diakonov,  {\em Towards a Chiral Theory of Nucleons}, in {\em Skyrmions and Anomalies}, World Scientific, 1987, p.27,
eds. M.~Jeżabek and M.~Praszałowicz.

\bibitem{Diakonov:1983hh}
D.~Diakonov and V.~Y.~Petrov,
{\em Instanton Based Vacuum from Feynman Variational Principle},
Nucl. Phys. B \textbf{245} (1984) 259,
doi:10.1016/0550-3213(84)90432-2.

\bibitem{Shuryak:1983ni}
E.~V.~Shuryak,
{\em Theory and phenomenology of the QCD vacuum},
Phys. Rept. \textbf{115} (1984) 151,
doi:10.1016/0370-1573(84)90037-1.

\bibitem{Scherer:2002tk}
S.~Scherer,
{\em Introduction to chiral perturbation theory},
Adv. Nucl. Phys. \textbf{27} (2003) 277,
[arXiv:hep-ph/0210398 [hep-ph]].



\bibitem{Diakonov:1983bny}
D.~Diakonov and M.~I.~Eides,
{\em Chiral Lagrangian from a functional integral over quarks},
JETP Lett. \textbf{38} (1983) 433.


\bibitem{Balog:1984upv}
J.~Balog,
{\em Effective Lagrangian from QCD Anomalies},
Phys. Lett. B \textbf{149} (1984) 197,
doi:10.1016/0370-2693(84)91583-1.

\bibitem{Praszalowicz:1989dh}
M.~Praszalowicz and G.~Valencia,
{\em Quark Models and Chiral Lagrangians},
Nucl. Phys. B \textbf{341} (1990) 27,
doi:10.1016/0550-3213(90)90261-B.

\bibitem{RuizArriola:1991gc}
E.~Ruiz Arriola,
{\em The Low-energy expansion of the generalized SU(3) NJL model},
Phys. Lett. B \textbf{253} (1991) 430,
doi:10.1016/0370-2693(91)91746-I.


\bibitem{Weinberg:1978kz}
S.~Weinberg,
{\em Phenomenological Lagrangians},
Physica A \textbf{96} (1979) 327,
doi:10.1016/0378-4371(79)90223-1.

\bibitem{Skyrme:1961vq}
T.~H.~R.~Skyrme,
{\em A Nonlinear field theory},
Proc. Roy. Soc. Lond. A \textbf{260} (1961) 127,
doi:10.1098/rspa.1961.0018.

\bibitem{Skyrme:1962vh}
T.~H.~R.~Skyrme,
{\em A Unified Field Theory of Mesons and Baryons},
Nucl. Phys. \textbf{31} (1962) 556,
doi:10.1016/0029-5582(62)90775-7.

\bibitem{Witten:1983tw}
E.~Witten,
{\em Global Aspects of Current Algebra},
Nucl. Phys. B \textbf{223} (1983) 422 ,
doi:10.1016/0550-3213(83)90063-9.

\bibitem{Witten:1983tx}
E.~Witten,
{\em Current Algebra, Baryons, and Quark Confinement},
Nucl. Phys. B \textbf{223} (1983) 433,
doi:10.1016/0550-3213(83)90064-0.

\bibitem{Witten:1979kh}
E.~Witten,
{\em Baryons in the 1/N Expansion},
Nucl. Phys. B \textbf{160} (1979) 57,
doi:10.1016/0550-3213(79)90232-3.

\bibitem{Goeke:2005fs}
K.~Goeke, J.~Ossmann, P.~Schweitzer and A.~Silva,
{\em Pion mass dependence of the nucleon mass and chiral
extrapolation of lattice data in the chiral quark soliton model},
Eur. Phys. J. A \textbf{27} (2006) 77,
doi:10.1140/epja/i2005-10229-5,
[arXiv:hep-lat/0505010 [hep-lat]].

\bibitem{Diakonov:1988mg}
D.~Diakonov, V.~Y.~Petrov and M.~Praszalowicz,
{\em Nucleon Mass and Nucleon $\Sigma$ Term},
Nucl. Phys. B \textbf{323} (1989) 53,
doi:10.1016/0550-3213(89)90587-7.

\bibitem{Guadagnini:1983uv}
E.~Guadagnini,
{\em Baryons as Solitons and Mass Formulae},
Nucl. Phys. B \textbf{236} (1984) 35,
doi:10.1016/0550-3213(84)90523-6.

\bibitem{Jain:1984gp}
S.~Jain and S.~R.~Wadia,
{\em Large-N baryons: Collective coordinates of the topological soliton in the SU(3) chiral model},
Nucl. Phys. B \textbf{258} (1985) 713,
doi:10.1016/0550-3213(85)90632-7.

\bibitem{Mazur:1984yf}
P.~O.~Mazur, M.~A.~Nowak and M.~Praszalowicz,
{\em SU(3) Extension of the Skyrme Model},
Phys. Lett. B \textbf{147} (1984) 137,
doi:10.1016/0370-2693(84)90608-7.

\bibitem{Chemtob:1985ar}
M.~Chemtob,
{\em Skyrme Model of Baryon Octet and Decuplet},
Nucl. Phys. B \textbf{256} (1985) 600,
doi:10.1016/0550-3213(85)90409-2.

\bibitem{Landau1977}
L.~D.~Landau and E.~M.~Lifshitz, 
{\em Quantum Mechanics: Non-Relativistic Theory}. Vol. 3 (3rd ed.) (1977), Pergamon Press. ISBN 978-0-08-020940-1.

\bibitem{Christov:1995vm}
C.~V.~Christov, A.~Blotz, H.~C.~Kim, P.~Pobylitsa, T.~Watabe, T.~Meissner, E.~Ruiz Arriola and K.~Goeke,
{\em Baryons as nontopological chiral solitons},
Prog. Part. Nucl. Phys. \textbf{37} (1996) 91,
doi:10.1016/0146-6410(96)00057-9.
[arXiv:hep-ph/9604441 [hep-ph]].

\bibitem{Adkins:1983ya}
G.~S.~Adkins, C.~R.~Nappi and E.~Witten,
{\em Static Properties of Nucleons in the Skyrme Model},
Nucl. Phys. B \textbf{228} (1983) 552,
doi:10.1016/0550-3213(83)90559-X.

\bibitem{Praszalowicz:1985bt}
M.~Praszalowicz,
{\em A Comment on the Phenomenology of the SU(3) Skyrme Model},
Phys. Lett. B \textbf{158} (1985) 264,
doi:10.1016/0370-2693(85)90968-2.


\bibitem {Yang:2016qdz}G.~S.~Yang, H.-Ch.~Kim, M.~V.~Polyakov and
M.~Praszalowicz,
{\em Pion mean fields and heavy baryons},
Phys.\ Rev.\ D \textbf{94} (2016)  071502,
doi:10.1103/PhysRevD.94.071502,
[arXiv:1607.07089 [hep-ph]].

\bibitem{Yang:2010fm}
G.~S.~Yang and H.~C.~Kim,
{\em Mass splittings of SU(3) baryons within a chiral soliton model},
Prog. Theor. Phys. \textbf{128} (2012) 397,
doi:10.1143/PTP.128.397,
[arXiv:1010.3792 [hep-ph]].


\bibitem{Blotz:1994wi}
A.~Blotz, M.~Praszalowicz and K.~Goeke,
{\em Axial properties of the nucleon with 1/N(c) corrections in the solitonic SU(3) NJL model},
Phys. Rev. D \textbf{53} (1996) 485,
doi:10.1103/PhysRevD.53.485,
[arXiv:hep-ph/9403314 [hep-ph]].

\bibitem{Walliser:2005pi}
H.~Walliser and H.~Weigel,
{\em Bound state versus collective coordinate approaches in chiral soliton models and the width of the Theta+ pentaquark},
Eur. Phys. J. A \textbf{26} (2005) 361,
doi:10.1140/epja/i2005-10180-5,
[arXiv:hep-ph/0510055 [hep-ph]].

\bibitem{Cheng:2006dk}
  H.~Y.~Cheng and C.~K.~Chua,
  {\em Strong Decays of Charmed Baryons in Heavy Hadron Chiral Perturbation Theory},
  Phys.\ Rev.\ D {\bf 75} (2007) 014006,
  doi:10.1103/PhysRevD.75.014006,
  [hep-ph/0610283].
  
\bibitem{Cheng:2015dk1}
  H.~Y.~Cheng and C.~K.~Chua,
  {\em Strong Decays of Charmed Baryons in Heavy Hadron Chiral Perturbation Theory: An Update},
  Phys.\ Rev.\ D {\bf 92} (2015)  074014,
  doi:10.1103/PhysRevD.92.074014,
  [arXiv:1508.05653 [hep-ph]].

\bibitem{Ellis:2004uz}
J.~R.~Ellis, M.~Karliner and M.~Praszalowicz,
{\em Chiral soliton predictions for exotic baryons},
JHEP \textbf{05} (2004) 002,
doi:10.1088/1126-6708/2004/05/002,
[arXiv:hep-ph/0401127 [hep-ph]].


\bibitem{deSwart:1963pdg}
J.~J.~de Swart,
{\em The Octet model and its Clebsch-Gordan coefficients},
Rev. Mod. Phys. \textbf{35} (1963) 916,
[erratum: Rev. Mod. Phys. \textbf{37}, 326-326 (1965)],
doi:10.1103/RevModPhys.35.916.

\bibitem{Biedenharn:1984qg}
L.~C.~Biedenharn, Y.~Dothan and A.~Stern,
{\em Baryons as Quarks in a Skyrmion Bubble},
Phys. Lett. B \textbf{146} (1984)  289, 
doi:10.1016/0370-2693(84)91698-8.

\bibitem{Biedenharn:1984su}
L.~C.~Biedenharn and Y.~Dothan,
{\em Monopolar Harmonics In SU(3)$_F$ as Eigenstates of the Skyrme-Witten Model for Baryons},
Print-84-1039 (DUKE).

\bibitem{Praszalowicz:2003ik}
M.~Praszalowicz,
{\em Pentaquark in the Skyrme model},
Phys. Lett. B \textbf{575} (2003) 234,
doi:10.1016/j.physletb.2003.09.049,
[arXiv:hep-ph/0308114 [hep-ph]].

\bibitem{Praszalowicz:1987em}
M.~Praszalowicz,
{\em SU(3) skyrmion},
Jagiellonian University preprint TPJU-5-87. In
{\em Skyrmions And Anomalies}
eds.
M. Jezabek and M. Praszalowicz, World Scientific 1987, p. 112.

\bibitem{Praszalowicz:1998jm}
   A.~Blotz, M.~Praszalowicz and K.~Goeke,
  {\em Rotational corrections to axial currents in semibosonized SU(3) Nambu-Jona-Lasinio model},
  Phys.\ Lett.\ B {\bf 317} (1993) 195,
  doi:10.1016/0370-2693(93)91592-B.
  [hep-ph/9308284].
  
  \bibitem{paradox}
  M.~Praszalowicz, T.~Watabe and K.~Goeke,
  {\em Quantization ambiguities of the SU(3) soliton},
  Nucl.\ Phys.\ A {\bf 647} (1999) 49,
  doi:10.1016/S0375-9474(99)00008-1,
  [hep-ph/9806431].

 \bibitem{Yang:2015era}
  G.~S.~Yang and H.-Ch.~Kim,
  {\em Hyperon Semileptonic decay constants with flavor SU(3) symmetry breaking},
  Phys.\ Rev.\ C {\bf 92} (2015) 035206,
  doi:10.1103/PhysRevC.92.035206,
  [arXiv:1504.04453 [hep-ph]].

\bibitem{Praszalowicz:2003tc}
M.~Praszalowicz,
{\em The Width of Theta+ for large Nc in chiral quark soliton model},
Phys. Lett. B \textbf{583} (2004) 96,
doi:10.1016/j.physletb.2003.12.056,
[arXiv:hep-ph/0311230 [hep-ph]].

\bibitem{Praszalowicz:2004dn}
M.~Praszalowicz,
{\em SU(3) breaking in decays of exotic baryons},
Acta Phys. Polon. B \textbf{35} (2004) 1625,
[arXiv:hep-ph/0402038 [hep-ph]].
 
\bibitem{Goeke:2009ae}
K.~Goeke, M.~V.~Polyakov and M.~Praszalowicz,
{\em On strange SU(3) partners of Theta+},
Acta Phys. Polon. B \textbf{42} (2011) 61,
doi:10.5506/APhysPolB.42.61
[arXiv:0912.0469 [hep-ph]].

\bibitem{Praszalowicz:2010me}
M.~Praszalowicz,
{\em Importance of Mixing for Exotic Baryons},
Acta Phys. Polon. B Supp. \textbf{3} (2010) 917,
[arXiv:1005.1007 [hep-ph]].

\bibitem{Jaffe:2004qj}
R.~L.~Jaffe,
{\em The width of the Theta+ exotic baryon in the chiral soliton model},
Eur. Phys. J. C \textbf{35} (2004) 221,
doi:10.1140/epjc/s2004-01815-4, 
[arXiv:hep-ph/0401187 [hep-ph]].

\bibitem{Diakonov:2004ai}
D.~Diakonov, V.~Petrov and M.~Polyakov,
{\em Comment on the Theta+ width and mass},
[arXiv:hep-ph/0404212 [hep-ph]].


\bibitem{Jaffe:2004dc}
R.~L.~Jaffe,
{\em Comment on hep-ph/0404212 by D. Diakonov, V. Petrov, and M. Polyakov},
[arXiv:hep-ph/0405268 [hep-ph]].

\bibitem{NA49:2003fxh}
C.~Alt \textit{et al.} [NA49],
{\em Observation of an exotic S = -2, Q = -2 baryon resonance in proton proton collisions at the CERN SPS},
Phys. Rev. Lett. \textbf{92} (2004) 042003,
doi:10.1103/PhysRevLett.92.042003,
[arXiv:hep-ex/0310014 [hep-ex]].

\bibitem{NA61SHINE:2020mti}
A.~Aduszkiewicz \textit{et al.} [NA61/SHINE],
{\em Search for an Exotic $S=-2, Q=-2$ baryon resonance in proton-proton interactions at $\sqrt{s_{NN}}$ = 17.3 GeV},
Phys. Rev. D \textbf{101}  (2020) 051101,
doi:10.1103/PhysRevD.101.051101,
[arXiv:1912.12198 [hep-ex]].



\bibitem{Kuznetsov:2008ii}
V.~Kuznetsov and M.~V.~Polyakov,
{\em New Narrow Nucleon N*(1685)},
JETP Lett. \textbf{88} (2008) 347,
doi:10.1134/S002136400818001X,
[arXiv:0807.3217 [hep-ph]].

\bibitem{GRAAL:2004ndn}
V.~Kuznetsov \textit{et al.} [GRAAL],
{\em $\eta$ photoproduction off the neutron at GRAAL: Evidence for a resonant structure at W = 1.67-GeV},
 Proceedings, 4th International Workshop on the Physics of Excited Nucleons (NSTAR 2004) ,
doi:10.1142/9789812702272\_0022,
[arXiv:hep-ex/0409032 [hep-ex]].

\bibitem{GRAAL:2006gzl}
V.~Kuznetsov \textit{et al.} [GRAAL],
{\em Evidence for a narrow structure at W\textasciitilde{}1.68-GeV in eta photoproduction on the neutron},
Phys. Lett. B \textbf{647} (2007) 23,
doi:10.1016/j.physletb.2007.01.041,
[arXiv:hep-ex/0606065 [hep-ex]].

\bibitem{Kuznetsov:2008hj}
V.~Kuznetsov, M.~V.~Polyakov, T.~Boiko, J.~Jang, A.~Kim, W.~Kim, H.~S.~Lee, A.~Ni and G.~S.~Yang,
{\em Evidence for a narrow N*(1685) resonance in eta photoproduction off the nucleon},
Acta Phys. Polon. B \textbf{39} (2008) 1949,
[arXiv:0807.2316 [hep-ex]].

\bibitem{CBELSA:2008epm}
I.~Jaegle \textit{et al.} [CBELSA and TAPS],
{\em Quasi-free photoproduction of eta-mesons of the neutron},
Phys. Rev. Lett. \textbf{100} (2008) 252002,
doi:10.1103/PhysRevLett.100.252002,
[arXiv:0804.4841 [nucl-ex]].

\bibitem{Shimizu:2008cwq}
H.~Shimizu, F.~Miyahara, T.~Nakabayashi, H.~Fukasawa, R.~Hashimoto, T.~Ishikawa, T.~Iwata, H.~Kanda, J.~Kasagi and T.~Kinoshita, \textit{et al.}
{\em N* (1670) observed at LNS, Sendai},
Contribution to NSTAR 2007, 
doi:10.1007/978-3-540-85144-8\_12.

\bibitem{Polyakov:2003dx}
M.~V.~Polyakov and A.~Rathke,
{\em On photoexcitation of baryon anti-decuplet},
Eur. Phys. J. A \textbf{18} (2003) 691,
doi:10.1140/epja/i2003-10029-y,
[arXiv:hep-ph/0303138 [hep-ph]].

\bibitem{Yang:2013tka}
G.~S.~Yang and H.~C.~Kim,
{\em Theta+ baryon, N*(1685) resonance, and pi N sigma term reexamined within the framework of a chiral soliton model},
PTEP \textbf{2013} (2013)  013D01,
doi:10.1093/ptep/pts044.


\bibitem{Shklyar:2006xw}
V.~Shklyar, H.~Lenske and U.~Mosel,
{\em eta-photoproduction in the resonance energy region},
Phys. Lett. B \textbf{650} (2007) 172,
doi:10.1016/j.physletb.2007.05.005,
[arXiv:nucl-th/0611036 [nucl-th]].

\bibitem{Anisovich:2008wd}
A.~V.~Anisovich, I.~Jaegle, E.~Klempt, B.~Krusche, V.~A.~Nikonov, A.~V.~Sarantsev and U.~Thoma,
{\em Photoproduction of eta mesons off neutrons from a deuteron target},
Eur. Phys. J. A \textbf{41} (2009) 13,
doi:10.1140/epja/i2009-10766-9,
[arXiv:0809.3340 [hep-ph]].

\bibitem{Doring:2009qr}
M.~Doring and K.~Nakayama,
{\em On the cross section ratio sigma(n)/sigma(p) in eta photoproduction},
Phys. Lett. B \textbf{683} (2010) 145,
doi:10.1016/j.physletb.2009.12.029,
[arXiv:0909.3538 [nucl-th]].

\bibitem {str}
R.~A.~Arndt, Y.~I.~Azimov, M.~V.~Polyakov, I.~I.~Strakovsky and R.~L.~Workman,
{\em Nonstrange and other unitarity partners of the exotic Theta+ baryon},
Phys. Rev. C \textbf{69} (2004) 035208,
doi:10.1103/PhysRevC.69.035208,
[arXiv:nucl-th/0312126 [nucl-th]].

\bibitem{Azimov:2006he}
Y.~I.~Azimov, V.~Kuznetsov, M.~V.~Polyakov and I.~Strakovsky,
{\em K*-couplings for the antidecuplet excitation},
Phys. Rev. D \textbf{75} (2007), 054014
doi:10.1103/PhysRevD.75.054014
[arXiv:hep-ph/0611238 [hep-ph]].

\bibitem{CLAS:2006anj}
R.~De Vita \textit{et al.} [CLAS],
{\em Search for the $\Theta^+$ pentaquark in the reactions gamma p ---\ensuremath{>} anti-K0 K+n and gamma p ---\ensuremath{>} anti-K0 K0p},
Phys. Rev. D \textbf{74}  (2006) 032001,
doi:10.1103/PhysRevD.74.032001,
[arXiv:hep-ex/0606062 [hep-ex]].

\bibitem{CLAS:2003wfm}
V.~Kubarovsky \textit{et al.} [CLAS],
{\em Observation of an exotic baryon with S = +1 in photoproduction from the proton},
Phys. Rev. Lett. \textbf{92} (2004) 032001,
[erratum: Phys. Rev. Lett. \textbf{92}(2004) 049902],
doi:10.1103/PhysRevLett.92.032001,
[arXiv:hep-ex/0311046 [hep-ex]].


\bibitem{CLAS:2006czw}
B.~McKinnon \textit{et al.} [CLAS],
{\em Search for the $\Theta^+$ pentaquark in the reaction gamma d ---\ensuremath{>} p K- K+ n},
Phys. Rev. Lett. \textbf{96} (2006) 212001,
doi:10.1103/PhysRevLett.96.212001,
[arXiv:hep-ex/0603028 [hep-ex]].

\bibitem{CLAS:2003yuj}
S.~Stepanyan \textit{et al.} [CLAS],
{\em Observation of an exotic S = +1 baryon in exclusive photoproduction from the deuteron},
Phys. Rev. Lett. \textbf{91} (2003) 252001,
doi:10.1103/PhysRevLett.91.252001
[arXiv:hep-ex/0307018 [hep-ex]].

\bibitem{Nakano:2017zrr}
T.~Nakano [LEPS and LEPS2],
{\em Present and Future of LEPS and LEPS2},
JPS Conf. Proc. \textbf{17} (2017) 061002,
doi:10.7566/JPSCP.17.061002.

\bibitem{Nakano:2017fui}
T.~Nakano [LEPS and LEPS2],
{\em Recent Results from LEPS},
JPS Conf. Proc. \textbf{13} (2017)  010007,
doi:10.7566/JPSCP.13.010007.

\bibitem{Yosoi:2019mno}
M.~Yosoi [LEPS/LEPS2],
{\em Recent results from LEPS and status of LEPS2},
EPJ Web Conf. \textbf{199} (2019) 01020,
doi:10.1051/epjconf/201919901020

\bibitem{Nakanoprivate}
T. Nakano, private communication.

\bibitem{Ahn:2023hiu}
J.~K.~Ahn and S.~H.~Kim,
{\em Search for $\Theta ^+$ in $K^+d\rightarrow K^0pp$ reaction at J-PARC},
J. Korean Phys. Soc. \textbf{82} (2023) 579,
doi:10.1007/s40042-023-00709-w.

\bibitem{Amarian:2006xt}
M.~Amarian, D.~Diakonov and M.~V.~Polyakov,
{\em To see the exotic Theta+ baryon from interference},
Phys. Rev. D \textbf{78} (2008) 074003,
doi:10.1103/PhysRevD.78.074003,
[arXiv:hep-ph/0612150 [hep-ph]].

\bibitem{Amaryan:2011qc}
M.~J.~Amaryan, G.~Gavalian, C.~Nepali, M.~V.~Polyakov, Y.~Azimov, W.~J.~Briscoe, G.~E.~Dodge, C.~E.~Hyde, F.~Klein and V.~Kuznetsov, \textit{et al.}
{\em Observation of a narrow structure in $p(\gamma,K_s)X$ via interference with $\phi$-meson production},
Phys. Rev. C \textbf{85} (2012) 035209,
doi:10.1103/PhysRevC.85.035209,
[arXiv:1110.3325 [hep-ex]].

\bibitem{CLAS:2012gcw}
M.~Anghinolfi \textit{et al.} [CLAS],
{\em Comment on the narrow structure reported by Amaryan et al},
[arXiv:1204.1105 [hep-ex]].

\bibitem{Workman:2022ynf}
R.~L.~Workman \textit{et al.} [Particle Data Group],
{\em Review of Particle Physics},
PTEP \textbf{2022}, 083C01 (2022),
doi:10.1093/ptep/ptac097.

\bibitem{Sekihara:2019cot}
T.~Sekihara, H.~C.~Kim and A.~Hosaka,
{\em Feasibility study of the $K^{+} d \to K^{0} p p$ reaction for the ''$\Theta ^{+}$'' pentaquark},
PTEP \textbf{2020} (2020) 063D03,
doi:10.1093/ptep/ptaa070,
[arXiv:1910.09252 [hep-ph]].

\bibitem{KLF:2020gai}
M.~Amaryan \textit{et al.} [KLF],
{\em Strange Hadron Spectroscopy with Secondary KL Beam in Hall D},
[arXiv:2008.08215 [nucl-ex]].

\bibitem{Amaryan:2024koq}
M.~J.~Amaryan {\em et al.} [KLF],
{\em Search for} $\Theta^+$ {\em in} $K_L p \rightarrow K^+ n$ {\em  reaction in KLF at JLab},
[arXiv:2401.05887 [hep-ex]].

\end{thebibliography}
\end{document}